\documentclass{article}
\usepackage{epsf}
\makeatletter
\newif\ifpr@pstyle \pr@pstylefalse
\newif\ifnons@qeq  \nons@qeqfalse
\newfont{\fourteencp}{cmcsc10 scaled\magstep2}
\newfont{\titlefont}{cmbx10 scaled\magstep2}
\newfont{\authorfont}{cmcsc10 scaled\magstep1}
\newfont{\fourteenmib}{cmmib10 scaled\magstep2}
	\skewchar\fourteenmib='177
\newfont{\elevenmib}{cmmib10 scaled\magstephalf}
	\skewchar\elevenmib='177
\newfont{\ninemib}{cmmib9} \skewchar\ninemib='177
\makeatletter
\newcommand\nonsequentialeqnum{
        \nons@qeqtrue
	\@addtoreset{equation}{section}
	\def\theequation{\arabic{section}.\arabic{equation}}}
\newif\ifp@bblock  \p@bblocktrue
\newcommand\nopubblock{\p@bblockfalse}
\newcommand\topspace{\hrule height 0pt depth 0pt \vskip}
\newcommand\p@bblock{\begingroup \tabskip=\hsize minus \hsize
	\baselineskip=1.5\ht\strutbox \topspace-2\baselineskip
	\halign to\hsize{\strut ##\hfil\tabskip=0pt\crcr
	\the\Pubnum\crcr\the\date\crcr}\endgroup}
\newcommand\YUKAWAmark{\hbox{
        \ifpr@pstyle\ninemib\else\elevenmib\fi
        Yukawa\hskip1mm Institute\hskip1mm Kyoto \hfill}}
\newtoks\date
\newtoks\Pubnum

\Pubnum={YITP-99-51}
\date={August 1999}
\newcommand{\frontpageskip}{\vspace{12pt plus .5fil minus 2pt}}
\def\@authoraddress{} \def\@title{}
\def\title#1{\gdef\@title{\frontpageskip
	\begin{center}{\titlefont #1}\end{center}\par}}
\def\@author#1{\frontpageskip\par\begin{center}{\authorfont #1}
	\end{center}
	\nobreak}
\def\author#1{\expandafter\def\expandafter\@authoraddress\expandafter
    {\@authoraddress{\@author{#1}}}}
\def\andauthor#1{\expandafter\def\expandafter\@authoraddress\expandafter
    {\@authoraddress{\frontpageskip\centerline{and}\@author{#1}}}}
\def\authors#1{\expandafter\def\expandafter\@authoraddress\expandafter
    {\@authoraddress{\frontpageskip\noindent #1}}}
\def\@address#1{\par\begin{center}{\sl #1}\end{center}\par}
\def\address#1{\expandafter\def\expandafter\@authoraddress\expandafter
    {\@authoraddress{\@address{#1}}}}
\def\andaddress#1{\expandafter\def\expandafter%
    \@authoraddress\expandafter
    {\@authoraddress{\par\centerline{\sl and}\@address{#1}}}}
\renewcommand{\thanks}[1]{\footnote{#1}}
\newlength{\paperbaselineskip}
\def\maketitle{\par
  \begingroup
       \def\thefootnote{\fnsymbol{footnote}}
	\thispagestyle{empty}
        \baselineskip=\paperbaselineskip
	\@maketitle
	\endgroup
	\setcounter{footnote}{0}
	\let\maketitle\relax \let\@maketitle\relax
	\let\@thanks\relax \let\@title\relax
	\let\@title\relax \let\@authoraddress\relax
	\let\thanks\relax}
\def\@maketitle{%
        \ifpr@pstyle\vspace{-1.0cm}\else\vspace{-1.7cm}\fi
	\YUKAWAmark\vskip0.6cm
	\ifp@bblock\p@bblock \else\hrule height 0pt \relax \fi
	\@title
	\@authoraddress
	}
\renewcommand{\abstract}{\par\frontpageskip\centerline{
             \ifpr@pstyle\twelvecp\else\fourteencp\fi Abstract}
	\vspace{8pt plus 3pt minus 3pt}}
\makeatother

\newcommand{\be}{\begin{equation}}
\newcommand{\ee}{\end{equation}}
\newcommand{\bea}{\begin{eqnarray}}
\newcommand{\eea}{\end{eqnarray}}
\begin{document}

\title{
	Spin Alignments in Heavy-ion Resonances
	\footnote{Talk given at CLUSTER '99, June 14-19, Rab, Croatia.}
}

\author{E. Uegaki}
\address{Physics Division, Department of Mechanical Engineering, 
	Akita University, Akita 010-8502, Japan\\
	E-mail: ue@uws13.phys.akita-u.ac.jp}

\andauthor{Y. Abe}
\address{Yukawa Institute for Theoretical Physics, Kyoto University, 
	Kyoto 606-8502, Japan\\
	E-mail: abey@yukawa.kyoto-u.ac.jp}  


\maketitle

\vspace{2.5truecm}

\begin{abstract}
{Recent particle-particle-$\gamma$ coincident measurements on a
$^{28}\rm Si+{}^{28}Si$ resonance have suggested ''vanishing spin alignments''.
New analyses for spin alignments by using a molecular model are reported.
Different aspects between di-nuclear systems with prolate deformed nuclei 
and those with oblate deformed nuclei are clarified.}
\end{abstract}

\vspace{\fill}
\pagebreak

\section{Introduction}

Resonances observed in heavy-ion scattering have offered intriguing subjects
in nuclear physics.  
$^{24}\rm Mg+{}^{24}Mg$ and $^{28}\rm Si+{}^{28}Si$ resonances exhibit 
very narrow decay widths, 
with prominent peaks correlated among the elastic and inelastic channels, 
which suggest rather long lived compound nuclear systems.\cite{Be}
>From the viewpoint of di-nuclear molecules, the present authors have studied 
normal modes around the equilibrium configurations, which are expected 
to be responsible to the observed resonances.\cite{Ue89,Ue94}

Very recently, $^{28}\rm Si+{}^{28}Si$ scattering experiment has been done 
on the resonance at $E_{\rm CM}=55.8$MeV at IReS Strasbourg.\cite{Nou}
Figure~1 shows angular distributions for the elastic and inelastic channels
$2^+$, $(2^+,2^+)$, respectively. The oscillating patterns are found to be 
in good agreement with $L=38$, which suggests $L=J=38$ dominance in the 
resonance, namely, {\it misalignments} of the fragment spins.
Angular distributions of $\gamma$-rays from $^{28}\rm Si$ first excited state 
to the ground state have been also measured in coincidence with 
two  $^{28}\rm Si$ nuclei detected at $\theta_{\rm CM} = 90^\circ$.
Figure~2 displays $\gamma$-ray intensities in the three panels,
where quantizations are made 
for $z$-axis to be parallel to the beam direction in (a),
$z$-axis parallel to normal to the scattering plane in (b) and
$z$-axis parallel to the fragment direction in (c).
The angular distribution in (b) shows characteristic "m=0" pattern, which 
suggests fragment spins $\bf I_1$ and $\bf I_2$ are on the scattering plane 
and is consistent with the misalignments observed in 
$^{28}\rm Si$ angular distributions.
Those features are much different from $^{12}\rm C+{}^{12}C$ and 
$^{24}\rm Mg+{}^{24}Mg$ systems, which exhibit spin alignments.\cite{CL88}  
The aim of the present paper is to clarify the mechanism
of the appearance of spin {\it misalignments} in the resonance of
$^{28}\rm Si+{}^{28}Si$ system.

\begin{figure}[h]
\begin{minipage}[t]{7cm}
\begin{center}
\epsfysize=93mm
\epsfbox{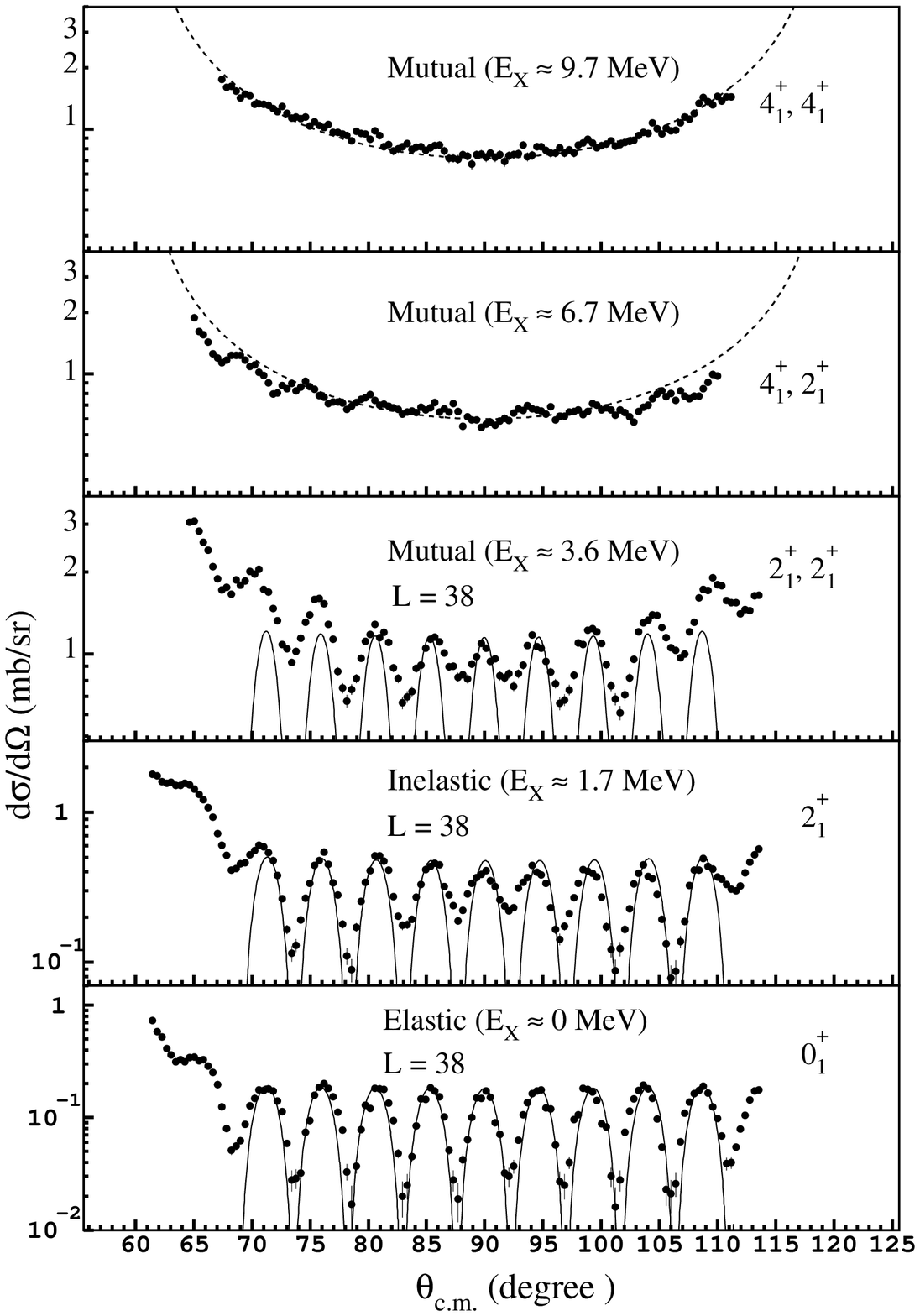} 
\caption{Experimental angular distributions for the elastic and 
inelastic scattering for $^{28}\rm Si+{}^{28}Si$ at $E_{\rm CM}=55.8$MeV. 
Solid curves show L=38 Legendre fits for comparison.}
\end{center}
\end{minipage}
\hspace*{1.0truecm}
\begin{minipage}[t]{7cm}
\begin{center}
\epsfysize=93mm
\epsfbox{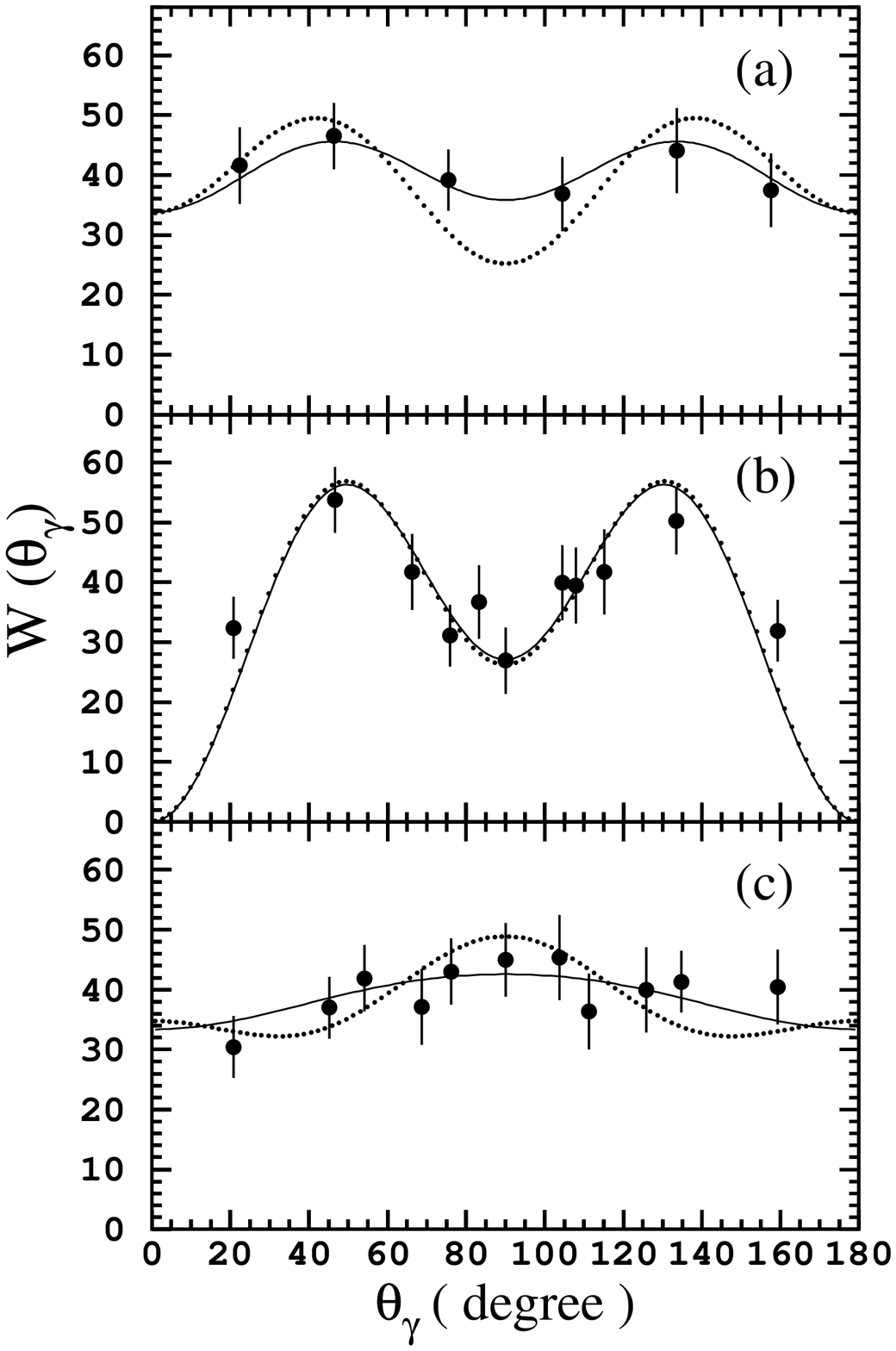} 
\caption{$\gamma$-rays distributions obtained from 
particle-particle-$\gamma$ coincident measurements.
Solid curves show $\chi^2$-fits. Dotted lines show the theoretical prediction
obtained by the molecular model with tilting mode (K-mixings). }
\end{center}
\end{minipage}
\end{figure}

\section{Di-nuclear Structure of $\bf {}^{28}Si+{}^{28}Si$ System} 
  
First, structures of the resonance states of $^{28}\rm Si+{}^{28}Si$ system 
and their normal modes around a stable configuration are briefly revisited. 
For simplicity, assuming a constant deformation 
and axial symmetry of the constituent nuclei, we have seven degrees of freedom 
$(q_{i})= (\theta _{1},\theta _{2},\theta _{3}, R,\alpha,\beta_1,\beta_2)$,
as illustrated in Fig.~3(a).
Consistently with the coordinate system, 
we introduce a rotation-vibration type wave function as basis one, 

\be
\Psi_\lambda \sim D_{MK}^J (\theta_i) 
                       \chi_{K}(R, \alpha,\beta_1,\beta_2), 
\label{eq:wf}
\ee
where  $\chi_{K}$ describes internal motions.
Dynamics of the internal motions have been solved around the equilibrium
and various normal modes such as butterfly vibrations have been 
obtained,\cite{Ue94} as is shown in Fig.~4(a).

Each mode has a characteristic feature with respect to the spin alignments.
For example, butterfly motion shows the total intrinsic spin $I=0$ dominance, 
namely, anti-alignment,
which is very consistent with the fragments angular distributions.\cite{Ue98}
However the indication by the $\gamma$-ray measurements that the spin vectors 
${\bf I}_1$ and $\bf I_2$ are both on the scattering plane 
provides more detailed information. 
In the following, we introduce a new mode 
in order to explain $\gamma$-rays data suggesting the fragments spin 
components ''$m=0$''.

\begin{figure}[h]
\leavevmode{
\hbox{\epsfysize=13pc \epsfbox{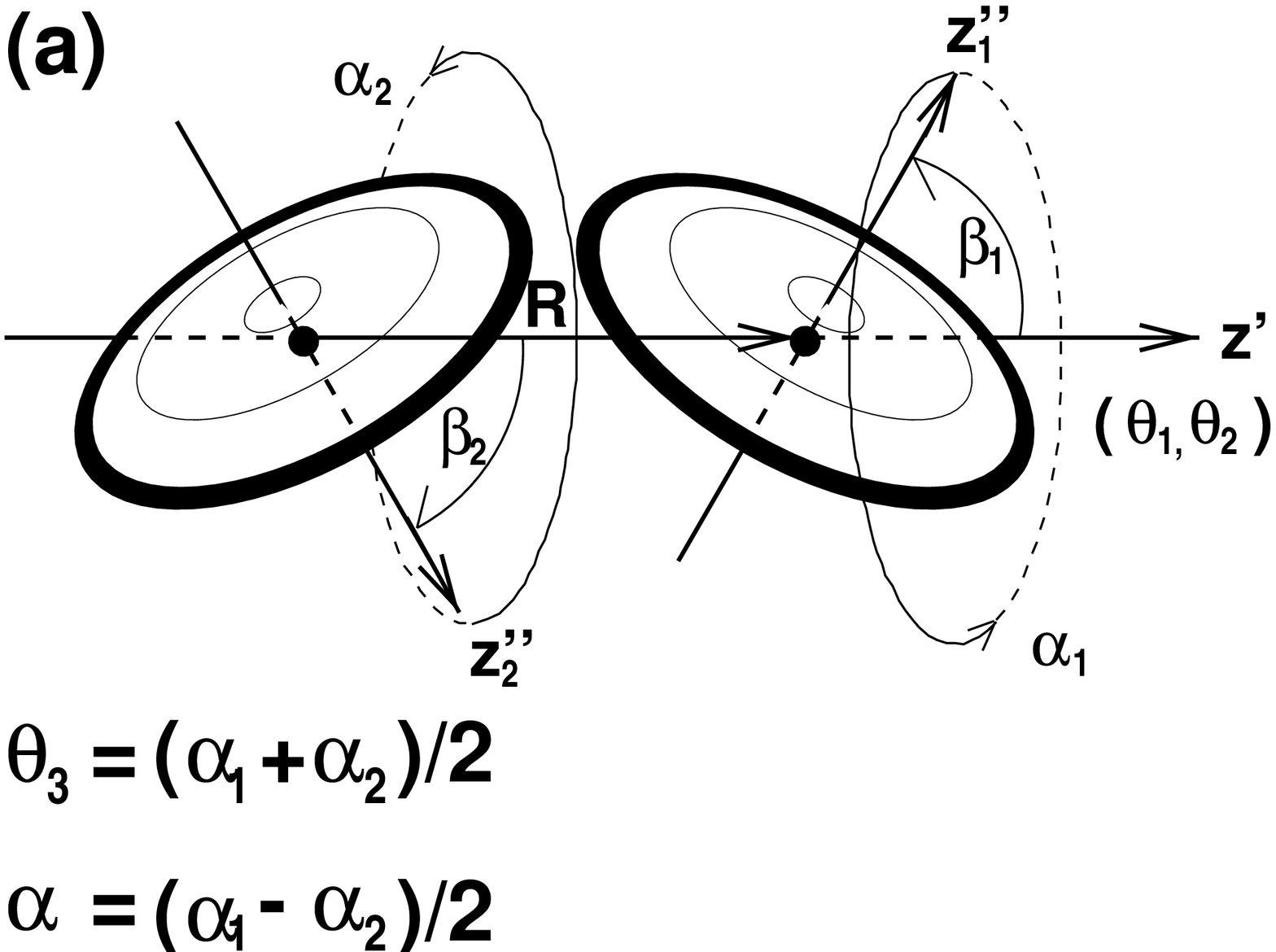}} 
{\hskip 0.8cm}
\hbox{\epsfysize=13pc \epsfbox{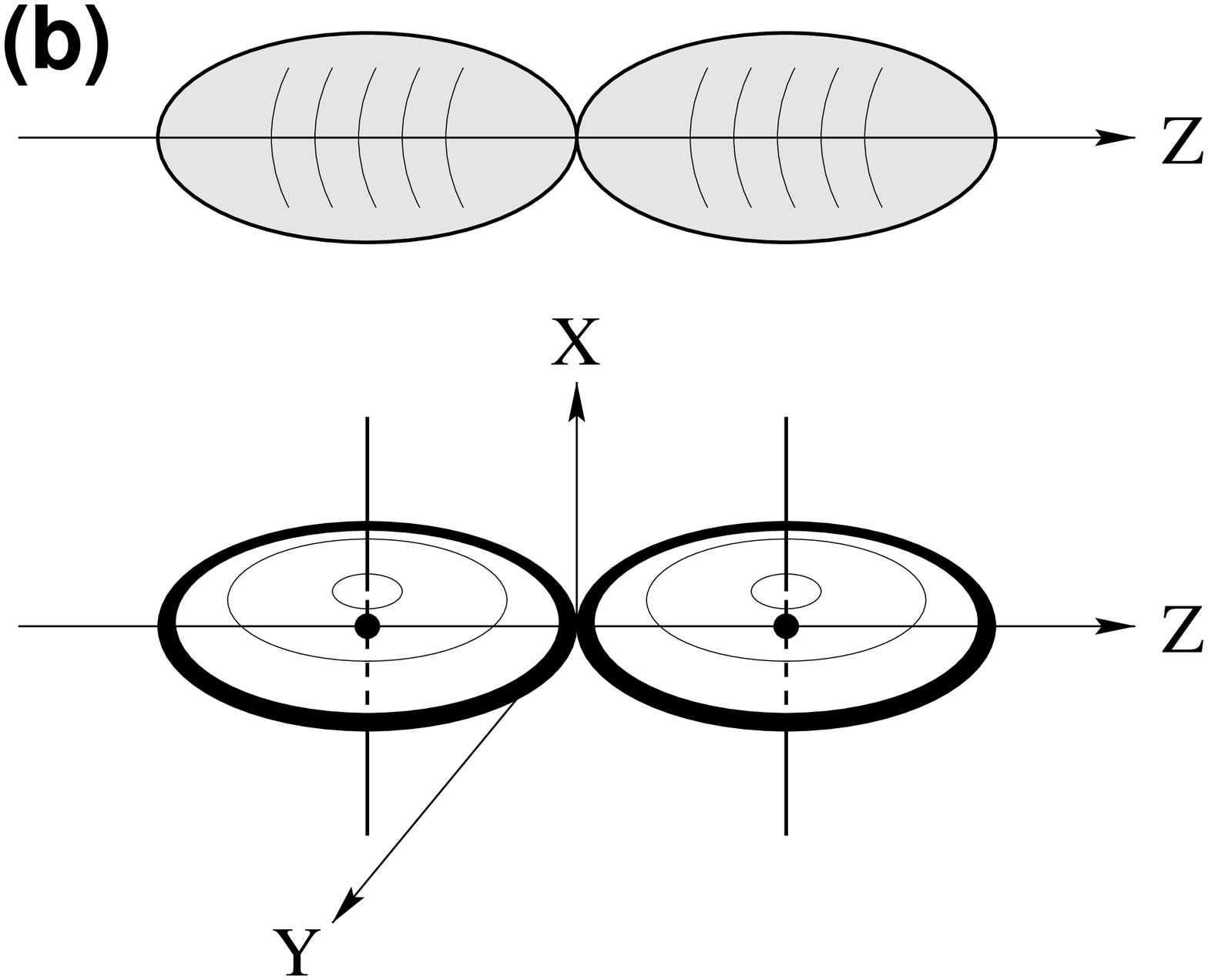}} 
}
\caption{(a) Coordinate system  of $^{28}\rm Si+{}^{28}Si$.
(b) Equilibrium configurations of di-nuclear systems. Upper panel is for 
$^{24}\rm Mg+{}^{24}Mg$ and lower panel for $^{28}\rm Si+{}^{28}Si$.}
\end{figure}

With extremely high spins such as $30 \sim 40\hbar$, stable di-nuclear
configurations tend to be ''elongated systems by the strong centrifugal force''.
In the prolate-prolate systems, a stable configuration is pole-to-pole one,
while in the oblate-oblate systems, it is an equator-to-equator one.
The former has axial symmetry as a whole, 
but the latter has axial asymmetry, as is displayed in Fig.~3(b).
{\it What would be expected from the difference of the symmetries?}
Approximately a triaxial system rotates around the axis with the maximum 
moments of inertia. In the oblate-oblate systems, thus two pancakes-like 
objects touching side-by-side as in the lower panel of Fig.~3(b) rotate 
around $x$-axis which is parallel to the normal to the reaction plane.  
The axial asymmetry, however, gives rise to a mixing of $K$-quantum 
numbers,\cite{Bo} which is different from the axial symmetric prolate-prolate 
cases.
In the high spin limit ($K/J  \sim 0$), the diagonalization in the $K$-space
is found to be equivalent to solving a differential equation of the harmonic
oscillator with parameters given by the moments of inertia.
Thereby, the solution is a gaussian, 
or a gaussian multiplied by an Hermite polynomial, 
\be
  f_n(K) = H_n({ K \over b } )
     \exp \biggl[ -{1 \over 2} \biggl( { K \over b } \biggr)^2  \biggr] ,
\label{eq:kf}
\ee
with the width of 
$ b= (2J^2  {I_K / \Delta } )^{1/4}$,
where
$I_K^{-1}=  I_z^{-1} - I_{\rm av}^{-1}$ 
and $ \Delta^{-1} =  I_y^{-1} - I_{\rm av}^{-1}$ 
with 
$I_{\rm av}^{-1}= {1 \over 2} ( I_x^{-1} + I_y^{-1})$.
The resultant energy spectrum is displayed in Fig.4(b) compared with 
the spectrum without $K$-mixings in Fig.4(a).
Consequences of the $K$-mixings to fragment spin orientations and then
to the $\gamma$-ray distributions are discussed in the next section.

\begin{figure}[t]
\epsfxsize=25pc 
\centerline{\epsfbox{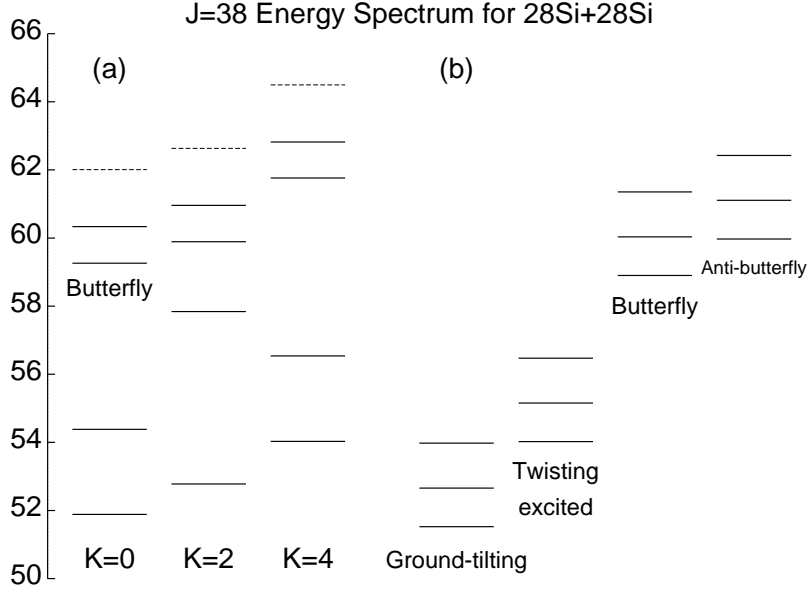}} 
\caption{Energy spectra for $^{28}\rm Si+{}^{28}Si$. (a) Without K-mixing 
and (b) with K-mixing. }
\end{figure}

\section{Spin Alignments}

We define scattering waves and the collision matrix $U_{c'c}$  such as

\be
\psi \sim (G_c -iF_c) - \sum_{c'} U_{c'c} (G_{c'} + iF_{c'}).
\label{eq:scat}
\ee

By using the $R$-matrix formula with one level approximation, 
we obtain \\
$U_{c' c} = e^{-i\phi'_{l'}}(-2\sqrt{2P_{c' l'}} \gamma_{c' l'}
           \sqrt{2P_{c l}} \gamma_{c l}) e^{-i\phi_l}/ \Gamma_{\rm total}$
for the inelastic scattering, 
where the reduced widths 
$\gamma_{c l}$
are calculated from the model wave functions.

The scattering amplitudes with specified magnetic substates are given by
\bea
A_{m_1 m_2}({\bf k',k}) & = & {2\pi \over ik} 
                  \sum_{L',S',M}(22m_1m_2|S'M'_S)(L'S'm'M'_S|JM)
\nonumber \\[4pt]
           & & \times  i^{J-L'} e^{i(\sigma_J + \sigma'_{L'})}U^J_{L'S'}
                        Y^*_{JM}({\hat k})  Y_{L'm'}({\hat k}')      .
\label{eq:samp}
\eea
The transition amplitudes for the $\gamma$-ray emissions 
from the polarized nuclei are discussed by several authors
(see for examlpe Ref.~8).
Note that in the experiment, 
only one of two emitted photons is detected in most cases, even with EUROGAM.
Note also that the intensities of the detectors are averaged 
over the azimuthal angle $\phi_\gamma$ in the data shown in Fig.~2
and then are expressed as 
$W(\theta_\gamma) = \sum_m  P_m W_m(\theta_\gamma)$,
where $P_m$ denotes the probability in the $m$ magnetic substate.

Now we calculate $P_m$'s and $W(\theta_\gamma)$'s with the molecular model. 
Combining with the lowest state of the tilting mode in Eq.~(\ref{eq:kf}),
i.e., $ f_0 (K) \sim  \exp (-K^2 / 2b^2)$,
we introduce a refined wave function,
\be
\Psi^{JM}_\lambda \sim \sum_K  \exp (-K^2 / 2b^2)  D_{MK}^J (\theta_i) 
                      \chi_{K}(R, \alpha,\beta_1,\beta_2), 
\label{eq:nwf}
\ee
where in general, $\chi_K$ can be any excitation mode.
A simple choice is that all the internal motions are zero-point ones. 
In Fig.~2, theoretical results for $b=1.3$ are shown by dotted lines, 
which are seen to be in good agreement with the data,
in all the three axes (a), (b) and (c).  This value of $b$ is consistent 
with the di-nuclear configuration of $^{28}\rm Si+{}^{28}Si$ system. 

\section{Concluding Remarks}

Differences between the oblate-oblate system($^{28}\rm Si+{}^{28}Si$) 
and the prolate-prolate system ($^{24}\rm Mg+{}^{24}Mg$) has been manifested.
Corresponding to the axial asymmetry by the equilibrium shape of
$^{28}\rm Si+{}^{28}Si$ system, 
we have introduced $K$-mixings and have succeeded to explain the characteristic
features for the spin misalignments.
Those $K$-mixings, namely, the {\it tilting mode} would be
a new facet in nuclear resonance phenomena.
Further experimental study is strongly desired.

\section*{Acknowledgments}

The authors are grateful to Dr.'s R.~Nouicer and C.~Beck 
for the informations on the experimental data and stimulating discussion.

\end{document}